\documentclass[a4paper, traditabstract]{aa}
\usepackage{graphicx}
\usepackage{epsfig}
\usepackage{color}
\usepackage{txfonts}
\usepackage{natbib}
\bibpunct{(}{)}{;}{a}{}{,} 

\def\approxlt{\lower.2em\hbox{$\buildrel < \over \sim$}}
\def\approxgt{\lower.2em\hbox{$\buildrel > \over \sim$}}

\newcommand{\HI}{\mbox{H{\sc i}}}
\newcommand{\HII}{\mbox{H{\sc ii}}}
\newcommand{\HIbf}{\mbox{H\hspace{0.155 em}{\footnotesize \bf I}}}
\newcommand{\HIit}{\mbox{H\hspace{0.155 em}{\footnotesize \it I}}}

\def\gtrsim{\mathrel{\hbox{\rlap{\hbox{\lower4pt\hbox{$\sim$}}}\hbox{$>$}}}}
\newcommand{\kms}{\mbox{{\rm km}\,{\rm s}$^{-1}$}}

\def\lesssim{\mathrel{\hbox{\rlap{\hbox{\lower4pt\hbox{$\sim$}}}\hbox{$<$}}}}

\def\la{\mathrel{\hbox{\rlap{\hbox{\lower4pt\hbox{$\sim$}}}\hbox{$<$}}}}
\def\ga{\mathrel{\hbox{\rlap{\hbox{\lower4pt\hbox{$\sim$}}}\hbox{$>$}}}}
\newcommand{\nan}{Nan\c{c}ay}
\newcommand{\am}[2]{$#1'\,\hspace{-1.7mm}.\hspace{.0mm}#2$}

\begin{document}

\authorrunning{Lehnert, van Driel, \& Minchin}

\title{Can galaxy growth be sustained through H{\Large I}-rich minor mergers?}

\titlerunning{Growth through \HI-rich minor mergers}

\author{M. D. Lehnert\inst{1,2} \and W. van Driel\inst{3,4} \and
R. Minchin\inst{5}}

\institute{Sorbonne Universit\'es, UPMC Universit{\'e} Paris VI, Institut
d'Astrophysique de Paris, 75014 Paris, France
\and
CNRS, UMR 7095, Institut d'Astrophysique de Paris, 98 bis boulevard
Arago, 75014 Paris, France
\and
GEPI, Observatoire de Paris, UMR 8111, CNRS, Universit\'e Paris Diderot,
5 place Jules Janssen, 92190 Meudon, France
\and
Station de Radioastronomie de \nan, Observatoire de Paris, CNRS/INSU USR 704, 
Universit\'e d'Orl\'eans OSUC, route de Souesmes, 18330 \nan, France
\and
Arecibo Observatory, National Astronomy and Ionosphere Center, Arecibo,
PR 00612, USA}

\date{Received .../Accepted ...}

\abstract{Local galaxies with specific star-formation rates
(star-formation rate per unit mass; sSFR$\sim$0.2-10 Gyr$^{-1}$) as
high as distant galaxies (z$\approx$1-3), are very rich in \HI. Those
with low stellar masses, M$_{\star}$=10$^{8-9}$ M$_{\sun}$, for example,
have M$_{\rm HI}$/M$_{\star}\approx$5-30. Using continuity arguments of
Peng et al. (2014), whereby the specific merger rate is hypothesized
to be proportional to the specific star-formation rate, and \HI\
gas mass measurements for local galaxies with high sSFR, we estimate
that moderate mass galaxies, M$_{\star}$=10$^{9-10.5}$ M$_{\sun}$,
can acquire sufficient gas through minor mergers (stellar mass ratios
$\sim$4-100) to sustain their star formation rates at z$\sim$2. The
relative fraction of the gas accreted through minor mergers declines with
increasing stellar mass and for the most massive galaxies considered,
M$_{\star}$=10$^{10.5-11}$ M$_{\sun}$, this accretion rate is insufficient
to sustain their star formation. We checked our minor merger hypothesis
at z=0 using the same methodology but now with relations for local normal
galaxies and find that minor mergers cannot account for their specific
growth rates, in agreement with observations of \HI-rich satellites
around nearby spirals. We discuss a number of attractive features, like
a natural down-sizing effect, in using minor mergers with extended \HI\
disks to support star formation at high redshift. The answer to the
question posed by the title, ``Can galaxy growth be sustained through
\HI-rich minor mergers?'', is maybe, but only for relatively low mass
galaxies and at high redshift.}


\maketitle

\section{Introduction}\label{sec:intro}  

The importance of merging in the growth of galaxies is widely recognized.
However, within the last decade, simulations have indicated that
cosmological accretion of gas through the filamentary structures that
develop as the Universe expands, also played a significant, perhaps
dominant role in supplying gas to galaxies \citep{keres05}. The physical
processes that may regulate the accretion of gas into the halos and onto
galaxies proper are not well understood \citep[e.g.][]{mitchell14}. All
gas accretion is driven by the evolution of the underlying dark matter
and whether it is cosmological gas accretion or merging depends on how
strongly the dark matter and baryons are coupled as they are accreted
\citep{danovich15}.

In a stimulating analysis developed to constrain the growth of galaxies
over cosmic time, \citet{peng14} assume a continuity in the stellar
mass evolution of galaxies. Using the constancy of the faint-end
slope of the galaxy mass function \citep[e.g.][]{ilbert13} and the
(slightly) negative slope of the specific star-formation rate-stellar
mass (sSFR-M$_{\star}$) relationship with redshift, they argue that
minor mergers (stellar mass ratios 4-10) must be important for the
stellar mass growth of galaxies. Allowing galaxies, as the they grow,
to accrete stellar mass due to merging with smaller galaxies naturally
restricts the evolution of the faint-end slope of the stellar mass
function as the overall stellar mass density increases and gives rise
to a negative slope in the sSFR-M$_{\star}$ relationship because not
all stellar growth occurs via {\it in situ} star formation. Adopting a
logarithmic slope of $-$0.1 for the sSFR-M$_{\star}$ relation and a faint
end slope of $-$1.4, \citet{peng14} determined that the specific merger
(stellar) mass accretion rate as a function of the satellite mass, sMMR
= 0.1 sSFR. In other words, about 10\% of the specific stellar growth
per unit time of a galaxy is due to accreting stars through mergers,
in addition to the usual definition of the sSFR as the specific growth
rate due to on-going star formation. They estimate that the range in
stellar growth through mergers is about $\pm$0.5 dex. They emphasized
that this model applies only to galaxies with masses below the break
in the mass function \citep[$\sim$10$^{10.7}$ M$_{\sun}$ which does not
evolve significantly with redshift; ][]{ilbert13}. The merger rate then
becomes, R(M$_{\rm sat}$) = x $\cdot$ sMMR(M$_{\rm \star, sat}$) = 0.1$\cdot$
x $\cdot$sSFR, where x is the ratio of the stellar mass of the central
galaxy, M$_{\rm \star, cen}$, to the satellite, M$_{\rm \star, sat}$.

Having characterized the merger rate in this way, if we include the
gas content of the merging galaxies, in particular their \HI\ content
using the ratio, M$_{\rm HI}$/M$_{\star}$, how much mass would then be
accreted for given stellar accretion rates? This is the goal of the
paper, to estimate the gas accretion rate for gas-rich minor mergers
and compare it to the amount of gas necessary to sustain galaxy growth
at z$\approx$2 and 0.

\section{Methodology}\label{Sec:methods}  

Our methodology relies on two strong assumptions. The first is that the
relative \HI\ gas content (i.e., the M$_{\rm HI}$/M$_{\star}$ ratio) of
galaxies at z$\sim$2 is the same as for our local sample with extremely
high sSFR, which we selected to have stellar masses and sSFR similar
to those at z$\sim$1-3 (Appendix A).  This assumption implies that the
relative \HI\ gas content of galaxies is not a function of redshift,
but that it increases with sSFR. As direct observations of the \HI\
content of distant galaxies are not currently feasible, to test this
assumption we used CO line observations of molecular gas content as a
proxy for their total gas content.

Our CO observations of our local high-sSFR galaxy sample suggest (Lehnert
et al. 2015, in prep.) they have molecular gas fractions, f$_{\rm gas}$ =
M$_{\rm H_2}$/(M$_{\star}$ + M$_{\rm H_2}$), as high as those of galaxies
at high redshift, and far exceeding those of local galaxies with average
sSFR \citep{tacconi10}. This comparison suggests that distant high-sSFR
galaxies may also have substantial quantities of \HI.

However, the similarity of molecular gas fractions between our low
redshift sample and distant galaxies in itself is not evidence that the
two populations are similar in their overall gas content. In fact, in
the literature, there are many papers that suggest the co-moving \HI\
content of the universe may be constant with redshift or evolve only
weakly \citep[cf.][]{prochaska09, martin10, zafar13}. As we show in
Section 4, a strongly evolving ratio of the \HI\ to stellar mass ratio is
consistent with a non-evolving co-moving density of \HI. This is simply
due to the fact that the co-moving stellar density evolves very strongly
with redshift \citep[][]{ilbert13}.

The second assumption is based on the continuity approach to estimate
the growth of galaxies with cosmic time \citep{peng14} -- which simply
states that galaxies must evolve continuously from one epoch to the next.
Adopting their formalism for galaxy growth (see Introduction), we will
assume that galaxies grow 10\% of their stellar mass through minor
mergers, and it implies that the merger probability increases linearly
with merger mass ratio. Given the uncertainties inherent in estimating
merger rates at high redshift \citep[e.g.][]{bertone09, bluck12}, it is
not clear if this is what is observed. If the merger probability goes
as the stellar mass ratio of the merger, then the mass acquired per
logarithmic bin is constant. \citet{peng14} suggest this is consistent
with merger rate estimates at high redshift, after a detailed comparison
\citep[at least for massive galaxies; e.g. ][]{newman12}.

To estimate the amount of gas accreted through minor mergers using the
formalism of \citet{peng14} requires an estimate of the gas to stellar
mass ratio as a function of stellar mass. While measuring the total
H$_2$ mass of distant galaxies at z=1-3 is feasible, it is currently
impossible to measure their \HI\ content, which we can only estimate
through models or by analogy with local objects that may be similar
to distant galaxies. To this end, we conducted a program of sensitive,
single dish observations of almost 200 nearby galaxies with extremely
high sSFR, meaning as high as observed at z=1-3 \citep[][]{elbaz07,
daddi07}. We describe the selection process of these galaxies and the
\HI\ observations and results in Appendix A.

\begin{figure}
\includegraphics[width=8.5cm]{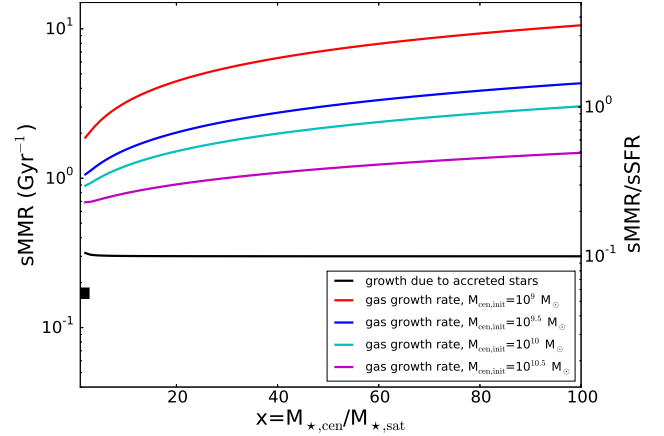}
\caption{The specific merger mass accretion rate (in units of Gyr$^{-1}$) 
as a function of the stellar mass ratio of mergers, x, for z=2. 
The lines represent an sMMR fixed at 10\% of the sSFR due to accreted stars 
(black solid line) and the total sMMR (including stars and gas) for 
various initial central galaxy stellar masses (indicated in the legend). 
We assumed sSFR = 3 Gyr$^{-1}$. On the right hand abscissa, we indicate the 
sSMR/sSFR ratio to compare the stellar mass growth due to merging and to 
star formation. The black point indicates the merger rate determined for 
z$\approx$2-4 by \citet{tasca14} for an assumed merger time scale of 
1 Gyr -- the value indicated is likely a lower limit since the time scale 
may be shorter.}
\label{fig:sMMRz2} 
\end{figure}

\section{Gas accretion through minor mergers}  

Assuming that the relative \HI\ content of galaxies at z$\sim$2
is the same as that of local galaxies with similarly high sSFR, we
can estimate the stellar plus gas accretion rate for high redshift
galaxies. To do this, we integrate the merger rate, R(M$_{\rm sat}$),
multiplied by the total baryonic mass, M$_{\star}$ + M$_{\rm HI}$ +
M$_{\rm H_2}$, over the range 1 to x$_{\rm lim}$, where x$_{\rm lim}$
spans the entire range of 1 to 100. We calculate these integrals as a
function of x$_{\rm lim}$ for a few fixed values of M$_{\rm cen}$ (log
M$_{\star}$ (M$_{\sun}$) = 9.0, 9.5, 10.0, 10.5) and have fixed the total
stellar mass accretion rate at 0.1 $\cdot$ sSFR for all ranges of the
integration with respect to x (Fig.~\ref{fig:sMMRz2}). To set the scale,
we assumed a constant sSFR=3 Gyr$^{-1}$, which is approximately the value
measured at z$\sim$2 \citep{daddi07} and similar to our sample of nearby
high-sSFR galaxies with measured \HI\ masses. For the contribution of
the \HI\ gas, we used the relationship log M$_{\rm HI}$/M$_{\star}$ =
$-$0.59 log M$_{\star}$ + 5.9 from our \HI\ survey of local galaxies
with high sSFR (Appendix A), which has been measured for objects with
8.0$\leq$log M$_{\star}$$\leq$10.8, or approximately over 0.002 to 1.2
times the break mass in the stellar mass function \citep{ilbert13}. As
already emphasized by \citet{peng14}, this is the mass range where the
stellar mass accretion model is appropriate.

For the contribution from H$_2$ we simply assumed a gas fraction of
0.5, consistent with CO observations of galaxies at z$\approx$2
\citep[][]{daddi10,aravena10,tacconi13} and of our sample of
nearby galaxies with comparably high sSFR and stellar masses
(Lehnert et al. 2015, in prep.). We note however that the exact
value of the molecular gas fraction assumed does not change the
result significantly as the \HI\ is the dominant contributor to the
gas accretion rate for the mass range probed in this analysis. If
the molecular gas fraction increases with decreasing stellar mass
\citep[e.g.][]{erb06,daddi10,tacconi10} then the gas accretion rate
through minor mergers would increase. Our analysis is conservative in
this respect.

Fig.~\ref{fig:sMMRz2} shows the total specific merger mass accretion over
a broad range of merger stellar mass ratios. The stellar mass accretion
rates assumed are consistent with observations \citep{tasca14}, although
we note that this comparison assumes a merger time of a Gyr, while it
is likely lower. We see that including the gas implies that galaxies
with relatively low initial stellar masses (M$_{\rm cen}$$<$10$^{10.5}$
M$_{\odot}$), the specific gas accretion rate through minor mergers, with
x$\ga$20-50, is sufficient to fuel the specific growth rate of galaxies
(as sMMR/sSFR$\ga$1). For high mass galaxies, near the break in the
mass distribution function, even minor mergers with relatively large mass
ratios, up to x=100, are insufficient to fully supply the necessary gas
to support the relative growth of galaxies. We further note including
even lower mass satellite galaxies (higher x) will not increase the gas
accretion rate substantially as although their \HI\ gas fractions are
high, their total \HI\ masses are small -- the effect saturates.

\subsection{Constraints from local galaxies}\label{subsec:localconstraints}  

Given that we do not have good constraints on the merger rate at high
redshifts, especially at low stellar mass ratios, and on the dependence
of the merger timescale on the mass ratio, it is important to test the
minor merger hypothesis locally. Analyses of observations of \HI\ in
low mass galaxies interacting/merging with nearby star-forming galaxies
suggest minor merger accretion rates of 0.1-0.3 M$_{\sun}$ yr$^{-1}$
\citep{sancisi08, diTeodoro14}.

\begin{figure}
\includegraphics[width=8.5cm]{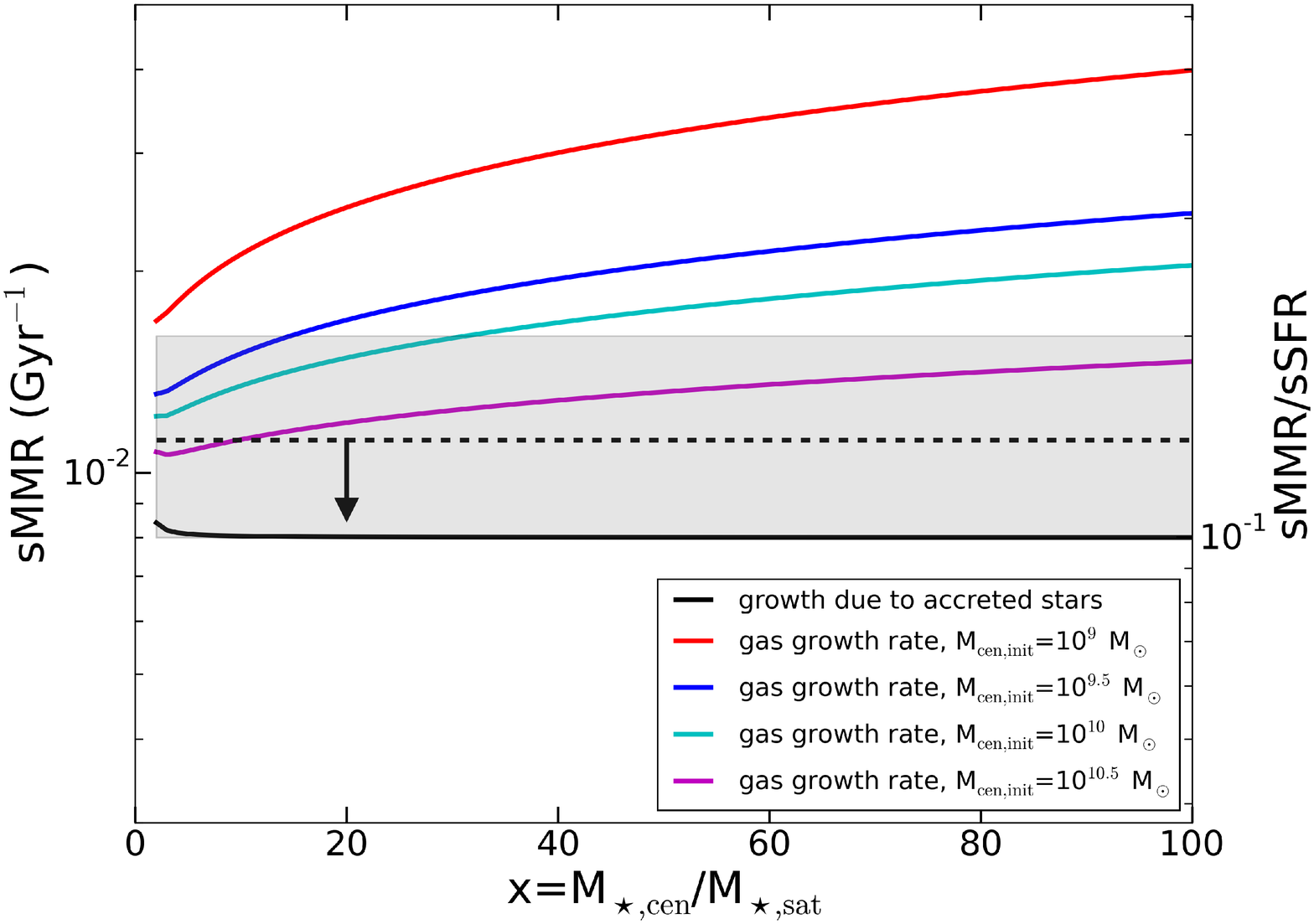}
\caption{The specific merger mass accretion rate (in units of Gyr$^{-1}$) 
as a function of the stellar
mass ratio of mergers, for z=0. The lines are exactly the same as in
Fig.~\ref{fig:sMMRz2} but now we use the measured mean \HI-to-stellar
mass ratio as a function of stellar mass from \citet{papastergis12}, and
sSFR=0.08 Gyr$^{-1}$. We also indicate the determinations of the \HI\
accretion rate from two surveys of local spirals \citep[][ grey shaded
region and dashed lines respectively; see text for details]{sancisi08,
diTeodoro14}.}

\label{fig:sMMRz0} 
\end{figure}

Using our model, but now assuming the measured mean relationship between
the \HI\ mass and stellar mass for local galaxies with typical sSFR
\citep[log M$_{\rm HI}$/M$_{\star}$ = $-$0.43 log M$_{\star}$ + 3.75;
][]{papastergis12} and a molecular gas fraction of 10\%, we again estimate
the stellar plus gas accretion rates as a function of stellar mass of the
central galaxy and the merger mass ratio, x (Fig.~\ref{fig:sMMRz0}). We
see that, in the local universe, minor mergers are far from being able
to supply the gas necessary to support the relative growth of galaxies,
and this, unlike at high redshift, even at the lowest initial central
galaxy stellar masses and for a limiting x=100.

Our predictions can be compared to the \HI\ accretion rates determined
for local galaxies \citep{sancisi08, diTeodoro14}, and are found to be
in general agreement. In the sample of \citet{diTeodoro14}, the average
stellar mass is 10$^{10.6}$ M$_{\sun}$ and their upper limit to the \HI\
mass accretion rate of 0.28 M$_{\sun}$ yr$^{-1}$ implies sMMR=0.007
Gyr$^{-1}$. However, it is not clear what the relative stellar masses
of their minor mergers are, but the baryonic mass ratios (stellar plus
\HI\ mass) are M$_{\rm bar, cen}$/M$_{\rm bar, sat}\ga$5. It is likely that the
stellar mass ratio is higher, since the satellites are richer in \HI\
than the central, more massive galaxy. In addition, the average sSFR of
their galaxies is less than what we assumed in our analysis. We have
therefore scaled up their sSFR by a factor of 1.6 for the purpose of
this comparison, to make it consistent with our assumptions about the
proportionality of the minor merger rate with the sSFR. Comparing with
the results from \citet{sancisi08} is more difficult due to a relative
lack of specific information about their sample and estimates. They
find an accretion rate of 0.1-0.2 M$_{\sun}$ yr$^{-1}$ for galaxies
with a star formation rate of $\approx$1 M$_{\sun}$ yr$^{-1}$ which
implies an sMMR/sSFR of 0.1-0.2. The latter is what we compare with in
Fig.~\ref{fig:sMMRz0}, not the sMMR directly.

We note that individual galaxies in our sample have uncertainties
of about 30\% in their estimated \HI\ to stellar mass ratios and that the
binned mass ratio data used in \citet{papastergis12} typically have about
the same uncertainties (see Appendix A). The uncertainty in the zero-point
of the fit for the high sSFR galaxies in the log M$_{\rm HI}$/M$_{\star}$
-- log M$_{\star}$ plane is 0.3 dex. If we take this into account,
analyzing it as a possible systematic uncertainty, implies that the
estimates of the sMMR as a function of x have a systematic range of about
$\pm$35\%. As we do not have an estimate of the uncertainties in the fits
estimated in \citet{papastergis12}, we will assume they are similar.
Such a level of uncertainty would not change our general conclusions of
whether or not minor mergers play a role at all and over what redshift range, 
although it would certainly change our detailed conclusions,
namely at what stellar masses are minor mergers likely to play a role in
supporting on-going star formation.

\subsection{Constraints from co-moving density of \HIit}\label{subsec:HIdensity}  

A basic test of whether or not distant galaxies are more \HI -rich than
local galaxies, as we have assumed, is to estimate the co-moving density
of \HI. To do this, we multiply the stellar mass function of galaxies
by the relationship between the \HI\ and stellar masses used in deriving
the specific merger gas accretion rate for distant and local galaxies.
Multiplying our best fit to the local galaxies with high sSFR (Appendix
A) by the stellar mass function at z=2.0-2.5, only considering the star
forming galaxies which dominate the mass budget \citep{ilbert13}, we
find that the co-moving density of \HI\ calculated in this way is only
a fraction, $\sim$20\%, of the total \HI\ co-moving density estimated
from sub- and damped Lyman-$\alpha$ absorbers \citep[sub-DLAs and -DLAs
respectively; ][]{zafar13}. The uncertainty in this estimate is large, at
least a factor of 2. If anything, since absorption line studies probe gas
that may not always be directly associated with the extended gas disks
of galaxies, we expect that using this method we will underestimate the
total co-moving \HI\ density. Furthermore, there are additional reasons
to not expect estimating the total co-moving mass of \HI\ as we have
done to agree with absorption line studies. \HI\ absorption lines probe
gas whose characteristics are likely different from emission line gas
(spin temperatures, dynamical state, fraction of warm to cold gas,
etc.) beyond whether or not the gas is predominately associated with
galaxy disks as we are assuming. We only intended this comparison to
gauge whether or not we are likely to over-estimate the \HI\ content of
distant galaxies, and observations suggest we are not.

\section{Discussion}  

Our analysis indicates that accreting gas through gas-rich minor mergers
provides a sufficient reservoir of gas to sustain the high specific star
formation rates observed for galaxies of relatively low mass (below about
M$^{\star}$) at high redshifts (z$\sim$2). There are several advantages
to this hypothesis: {\it (1)} it may mimic cosmological accretion; {\it
(2)} it will lead to shorter merger times than if we consider only their
stellar masses; {\it (3)} it will not thicken disks substantially; {\it
(4)} it may lead to relatively metal-rich galactic outskirts; and {\it
(5)} it is consistent with weak or no evolution in the co-moving density
of \HI. We very briefly discuss each of these possible advantages.

{{\it H}{\footnotesize \it I}\it-rich minor mergers may mimic cosmological
gas accretion:} If the \HI\ disks at high redshift are akin to their
low redshift counterparts, they will be rotationally supported and
dynamically cold, especially in their outer regions \citep[velocity
dispersions, $\sigma_{\rm HI}$$\sim$6-10 km s$^{-1}$ and v$_{\rm
rot}$/$\sigma_{\rm HI}$=3-10; ][]{stilp13a, stilp13b}. In addition,
the warm neutral gas has a very short turbulent dissipation timescale,
of-order 10 Myr, meaning, if disturbed, it will dissipate energy injected
as turbulence quickly -- in a dynamical time or less even for low mass
galaxies. In several important ways, the \HI\ gas in extended disks would
behave similarly to cold streams that have penetrated deeply into the
halo \citep{danovich15}. The small merging galaxies would be accreted
along filaments, their \HI\ gas is cold and stable, and they have high
angular momentum both from rotation and due to their orbits. In fact,
the extended \HI\ disks may induce a cooling wake in the hotter halo gas
as the galaxy falls deeper into the halo potential allowing more gas to
be accreted beyond that provided directly by the merger \citep{marasco12}
-- this would significantly increase the impact of direct gas accretion
through minor mergers.

{{\it H}{\footnotesize \it I}\it-rich minor mergers will have relatively
short(er) merging times:} In gas-rich mergers the dominating, dynamically
cold gaseous component would substantially decrease the merger time scale
\citep{lotz08, hopkins10}. Galaxies with stellar masses of 10$^{8-9}$
M$_{\sun}$ have approximately 10 times as much of their mass in \HI\ as
in stars. If regarded as truly minor mergers at 1:100, considering only
their stellar masses, they are actually $\sim$1:10 mergers considering
their gas content and will have a substantially reduced merger timescale
\citep[by a factor of 2 or more; ][]{hopkins10}.

{{\it H}{\footnotesize \it I}\it-rich minor mergers will not thicken
galactic disks substantially:} Minor mergers, especially frequent
ones as advocated in this analysis, might substantially heat the disk
and result in thick(er) disks. However, if the gas fractions are high
in the merging satellite, its impact on the thickening of the disk is
relatively small \citep{hopkinsP08}. While we do not show the heating
rate of the galaxies in our simple model, the impact on the vertical
heating rate of minor mergers is proportional to 1 - f$_{\rm gas,total}$,
the ratio of the total gas mass (\HI\ + H$_2$ + He) and the total mass
(gas + stars) of the galaxy. Since the majority of minor mergers are
gas rich, f$_{\rm gas,total}$ $>>$0.9, their impact on the heating the
disk of the central galaxy will be small \citep{hopkinsP08}.

{{\it H}{\scriptsize \it I}\it-rich minor mergers may lead to relatively
metal-rich galactic outskirts:} Minor mergers may also allow for
metal enriched gas to be accreted. Observations of \HII\ regions in the
outskirts of nearby galaxies suggest that they have attained substantial
metal enrichment \citep{bresolin12} which cannot be due to the currently
observed levels of star formation, even if sustained for a Hubble time
\citep{bresolin12}. While not demanding it, these results suggest gas
is accreted already enriched, as would be the case for low-mass satellite
galaxies.

{{\it H}{\scriptsize \it I} \it may be the conveyor belt sustaining the
growth of galaxies and growing galaxies through gas-rich minor mergers is
consistent with no evolution in the co-moving density of H{\scriptsize
\it I}:} This idea means that \HI\ is the ultimate gas reservoir which, 
through its own collapse or conversion to molecular gas (a conveyor), sustains the
long-term star formation and thus the growth of galaxies. If true, 
we would expect that as the stellar mass content of the
universe increased, the \HI\ to stellar mass ratio would decrease roughly in
lock step, and thus the total \HI\ content of the universe would
remain roughly constant. 
To test whether our scaling of the \HI\ to stellar mass ratios
of galaxies in the local and distant, z$\approx$2, universe, is consistent
with this conveyor belt hypothesis, we need to estimate, $\Omega_{\rm
HI}$, the co-moving density of \HI\ relative to the closure density.
$\Omega_{\rm HI}$ is constant over the redshift range 1-5, and maybe
even down to z=0 \citep{zafar13, martin10}, while the star formation
rate density has declined significantly and the co-moving stellar mass
density of the universe has increased \citep{ilbert13}. Although we
do not account for all of the \HI\ density estimated from sub- and
DLAs at high redshift \citep{zafar13} in our analysis, only $\sim$1/5
(Sect.~\ref{subsec:HIdensity}), we can estimate the evolution of the
co-moving density of \HI\ implied by our analysis to check for consistency
with this hypothesis. 

So is our analysis consistent with a constant amount of \HI\ per unit
co-moving volume, supporting the hypothesis that \HI\ is the reservoir
out of which star formation is sustained?  For this estimate, we assume
the mass functions given in \citet{ilbert13} for star forming galaxies at
z=2$-$2.5 and z=0.2$-$0.5. While they do not estimate the stellar mass
distribution function at z=0, the ultimate evolution of the total \HI\
content is likely small \citep{zafar13} as is the change in stellar
mass density. Doing this, using our ratio for \HI\ to stellar masses
as a function of stellar mass (Appendix A) for z=2$-$2.5 and that from
\citet{papastergis12} for galaxies at z=0, we find that the ratio of \HI\
densities, $\rho_{\rm HI}$, at z$\sim$2 and z$\sim$0 is about 1.5 with
an uncertainty of about a factor of $\sim$3. Even though the estimate is
crude, it does suggest that overall, our analysis is consistent with no
evolution in the co-moving density of \HI\ and implies that \HI\ gas is
the reservoir out of which galaxies move up the stellar mass hierarchy.

In this picture, this reservoir may be supplied through minor mergers,
at least at high redshifts and for relatively low mass galaxies.
Furthermore, downsizing is a natural result of galaxies becoming
increasingly gas starved as they grow in mass and, overall, as their
redshift decreases.

\begin{acknowledgements} 
We wish to thank the anonymous referee for carefully reading our
manuscript and making a number of extremely useful comments. We wish
to also thank Paola Di Matteo for her comments on and enthusiasm for
this work. The \nan\ Radio Telescope is operated as part of the Paris
Observatory, in association with the Centre National de la Recherche
Scientifique (CNRS) and partially supported by the R\'egion Centre in
France.  The Arecibo Observatory is operated by SRI International under a
cooperative agreement with the National Science Foundation (AST-1100968),
and in alliance with Ana G.  M\'endez-Universidad Metropolitana, and
the Universities Space Research Association.
\end{acknowledgements}

\bibstyle{aa}
\bibliographystyle{aa}

\bibliography{HImergersrefs}

\begin{appendix}
\section{Selection of a local high-sSFR galaxy sample, \HIbf\ observations and results}

\subsection{Sample selection}

All galaxies were selected using the Sloan Digital Sky Survey (SDSS)
Data Release DR7\footnote{http://www.sdss.org/dr7/}, in combination
with certain physical parameters of the galaxies (stellar masses,
star-formation rates, rest-frame optical colors, etc.) which were
taken from the publicly available SDSS ``value added'' MPA/JHU catalogs
\citep{brinchmann04, kauffmann03, salim07, tremonti04}. The consecutive
steps in the selection process are as follows.

\begin{enumerate}

\item {Redshift:} 0.01$<$z$<$0.05. We chose z=0.01 as our lower limit so
that the SDSS spectrograph fiber subtends at least the central 200 pc to
give meaningful results \citep[see][for this in comparison with distant
galaxies]{elbaz07}. The upper limit of z=0.05 (or a radial velocity of
15,000 \kms) is the estimated limit for a good \HI\ detection rate in 
our survey;

\item {Color:} rest-frame (g-r)$<$0.63 mag. We selected only the bluest
25\% of all galaxies in the MPA/JHU catalogs;

\item {sSFR:} highest 5\% of galaxies with blue rest-frame colors. 
We fitted a line to the resulting main sequence in the 
log(sSFR)--log(M$_{\star}$) plane, which has a slightly negative
slope, $-$0.2. We then, as a function of stellar mass, selected the
5\% of galaxies with the highest sSFRs. Although low mass galaxies on
average have a higher sSFR than the more massive ones, all objects in all
M$_{\star}$ ranges span similar ranges in sSFR as the distant galaxies
typically do: they have sSFR$\approx$0.2--10 Gyr$^{-1}$, compared to
0.7 Gyr$^{-1}$ and 2 Gyr$^{-1}$, for average main sequence galaxies at
z$\sim$1 and z$\sim$2, respectively \citep[e.g.][]{elbaz07, daddi07};

\item {Total stellar mass:} M$_{\star}\geq10^8 M_{\odot}$. We adopted
this limit in order to avoid galaxies in the low-mass range
where the SDSS is becoming highly incomplete. This lower limit in mass
also (generously) ensures that our sample spans the entire stellar mass
range of galaxies with well-determined sSFR at z$\sim$1-3.

\end{enumerate}

\subsection{H{\footnotesize \sl I} observations and results}

Our final high-sSFR local galaxy sample contains 2698 objects. Of those,
we have 329 clear \HI\ detections at our disposal for further analysis
(104 from our \nan\ and 23 from our Arecibo surveys, and 202 from the
literature), as well as 7 marginal detections (6 \nan\ and 1 Arecibo)
and 39 non-detections (29 \nan\ and 10 Arecibo). Publication of the
full results is in preparation.

The \nan\ Radio Telescope (NRT) is a 100 meter-class instrument. Its
HPBW is \am{3}{5} in right ascension, while in declination it is 23$'$
for $\delta$$<$20$^{\circ}$, increasing to 30$'$ at the northern limit
of our survey. Observations were taken in position-switching mode,
in the period January 2012 - March 2013 for a total of 400 hours of
telescope time. The data were reduced using standard NRT routines,
smoothing to a velocity resolution of 18~\kms.

The Arecibo radio telescope has a 305 m diameter. The HPBW of the L-band
wide (LBW) receiver used is \am{3}{1}$\times$\am{3}{5}. Observations
were taken in position-switching mode, in the period January 2013 - May
2014 for a total of 14 hours of telescope time. The data were reduced
using standard Arecibo Observatory IDL routines, smoothing to a velocity
resolution of 20~\kms.

We found published \HI\ detections of a further 293 target galaxies, of
which we used 202 for further analysis. They were found predominantly
(68\%) from matching our SDSS coordinates within a search radius of
4$''$ to those in the $\alpha$.40 catalog \citep{haynes11} of the
blind Arecibo ALFALFA survey, from which we used only the reliable
quality 1 detections, which have a signal-to-noise ratio greater than
6.5 \citep[see][]{saintonge07}. We also searched the on-line NED and
HyperLeda databases, within a radius of 15$''$. However, after visual
inspection of SDSS images, we noted that 54 detections are associated in
the MPA/JHU catalog with SDSS spectroscopic sources which are actually
sub-structures in the galaxies outside the nuclear region, such as star
forming regions. As their associated SDSS photometric sources do not
represent the entire galaxy, the estimated stellar masses given in the
MPA/JHU catalog are therefore likely (severe) underestimates for the
galaxy as a whole, and we did not use them further for our study. We
also discarded 37 cases where one or more other galaxies within the radio
telescope beam were likely to confuse the \HI\ detection of the target.

\subsection{Uncertainties}

\begin{figure}
\includegraphics[width=8.5cm]{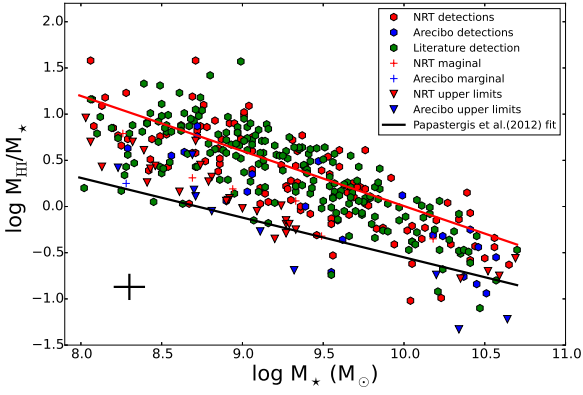}
\caption{The logarithm of the \HI\ to stellar mass ratio, log
M$_{HI}$/M$_{\star}$, as a function of the logarithm of the stellar
mass, log M$_{\star}$ (in units of M$_{\sun}$) for the galaxies in
our high sSFR sample. Detections with S/N$>$5 are shown as hexagons,
marginal detections of S/N 3-5 are shown as crosses, and upper limits
shown as downward pointing triangles. The colors indicate the telescope
used, or if the detection are from the literature as indicated in the
legend. We indicate the typical uncertainty in each quantity
by the black cross in the lower left corner of the plot (see text
for details). The red line is the best fit using the Buckley-James
method to fit censored data \citep[i.e., including our upper limits,
which gives, log M$_{HI}$/M$_{\star}$ = $-$0.59 log M$_{\star}$
+ 5.9;][]{isobe86}. The black line is the fit to a sample of local
galaxies with typical sSFR \citep{papastergis12}.}
\label{fig:massratios}
\end{figure}

The uncertainties in the stellar mass estimates of our sample of
galaxies were determined by taking the 1$\sigma$ confidence level of each
galaxy from the MPA/JHU value added catalogs. To estimate the typical
uncertainty, we simply calculate the average of all uncertainties. Showing
each individual uncertainty in Fig.~\ref{fig:massratios} would make
the plot illegible, however. We find that the average uncertainty
in the stellar mass estimates we used is 20\%. \citet{papastergis12}
used a somewhat different way to estimate stellar masses of their local
\HI-detected galaxies with typical sSFR, see \citet{huang12} for details.
As we do not have a simple way of estimating the systematic uncertainties
between the two stellar mass estimates we will ignore this, but we note
that in our experience different mass estimates usually agree within about
0.3 dex \citep[e.g.,][]{drory04, moustakas13, sorba15}. For both our NRT
and Arecibo \HI\ data, we find a typical relative uncertainty of about
15\% and a systematic uncertainty of about 10\%, which we determined
through regular observations of \HI\ line flux calibrator galaxies.
To estimate the total uncertainty we added each uncertainty in quadrature.
We have indicated the typical uncertainty in Fig.~\ref{fig:massratios}
for both the stellar mass and the ratio of \HI\ to stellar mass.

The uncertainties in our fit to the high sSFR galaxies are 0.03 in the slope
and 0.3 in the zero-point. Thus, the latter will effect our conclusions the 
most significantly. Unfortunately, \citet{papastergis12} do not provide an 
estimate of the uncertainties in their fit.

\end{appendix}

\end{document}